\documentclass{article}

\usepackage{arxiv}

\usepackage[utf8]{inputenc} 
\usepackage[T1]{fontenc}    
\usepackage{hyperref}       
\usepackage{url}            
\usepackage{booktabs}       
\usepackage{amsfonts}       
\usepackage{nicefrac}       
\usepackage{microtype}      
\usepackage{graphicx}
\usepackage{doi}
\usepackage{amsmath,amssymb,latexsym, amscd}
\usepackage{exscale, cite, eps fig, graphics}
\usepackage{array}
\usepackage{amsmath}
\usepackage{float}
\restylefloat{table}
\usepackage{placeins}
\usepackage{longtable}
\usepackage{makecell}
\usepackage[utf8]{inputenc}
\usepackage{graphicx}
\usepackage{caption}
\usepackage{subcaption}
\usepackage{environ}
\usepackage{mathtools}
\usepackage{booktabs}
\usepackage{multirow}
\usepackage{algorithmic}
\usepackage{graphicx}
\usepackage{textcomp}
\usepackage{xcolor}
\usepackage{amsthm}

\title{Tropical cyclone intensity estimations over the Indian ocean using Machine Learning}

\author{ \href{https://orcid.org/0000-0002-9818-8966}{\includegraphics[scale=0.06]{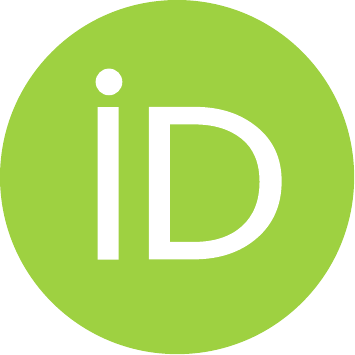}\hspace{1mm}Koushik Biswas} \\
	Department of Computer Science, IIIT Delhi\\
	New Delhi, India, 110020. \\
	\texttt{koushikb@iiitd.ac.in} \\
	\And
	\href{https://orcid.org/0000-0002-5464-929X}{\includegraphics[scale=0.06]{orcid.pdf}\hspace{1mm}Sandeep Kumar} \\
	Department of Computer Science, IIIT Delhi\\
	\&\\
	Department of Mathematics,\\ Shaheed Bhagat Singh College, University of Delhi\\
	New Delhi, India, 110020.\\
	\texttt{sandeepk@iiitd.ac.in, sandeep\_kumar@sbs.du.ac.in} \\
	\AND
	 Ashish Kumar Pandey\\
	Department of Mathematics, IIIT Delhi\\
	New Delhi, India, 110020.\\
	\texttt{ashish.pandey@iiitd.ac.in } \\
}

\begin{document}
\maketitle

\begin{abstract}
Tropical cyclones are one of the most powerful and destructive natural phenomena on earth. Tropical storms and heavy rains can cause floods, which lead to human lives and economic loss. Devastating winds accompanying cyclones heavily affect not only the coastal regions, even distant areas. Our study focuses on the intensity estimation, particularly cyclone grade and maximum sustained surface wind speed (MSWS) of a tropical cyclone over the North Indian Ocean. We use various machine learning algorithms to estimate cyclone grade and MSWS. We have used the basin of origin, date, time, latitude, longitude, estimated central pressure, and pressure drop as attributes of our models. We use multi-class classification models for the categorical outcome variable, cyclone grade, and regression models for MSWS as it is a continuous variable. Using the best track data of 28 years over the North Indian Ocean, we estimate grade with an accuracy of 88\% and MSWS with a root mean square error (RMSE) of 2.3. For higher grade categories (5-7), accuracy improves to an average of 98.84\%. We tested our model with two recent tropical cyclones in the North Indian Ocean, Vayu and Fani. For grade, we obtained an accuracy of 93.22\% and 95.23\% respectively, while for MSWS, we obtained RMSE of 2.2 and 3.4 and $R^2$ of 0.99 and 0.99, respectively.  
\end{abstract}


\section{Introduction}\label{sec:intro}

Tropical cyclones are rapidly rotating storm systems centered in a low-pressure region. Tropical cyclones cause heavy rain, strong wind, large storm surges near landfall, and tornadoes, which results in loss of property and lives. About 1.9 million people have died because of tropical cyclones worldwide during the last two centuries \cite{estimate2005robert,cropmanage}. The North Indian ocean (which includes the Bay of Bengal and Arabian sea) alone has seen some of the most devastating tropical cyclones. In 2019, both coasts of India experienced substantial damages because of Vayu and Fani.   

It is of high importance to estimate the intensity of a tropical cyclone. A standard indicator of the intensity of the storm is the maximum sustained surface wind speed (MSWS). The World Meteorological Organization categorizes the low-pressure systems using the ranges of  MSWS of the tropical cyclones \cite{mssw}. The categorization can be used to determine possible storm surges and damage impact on land \cite{Victor2009Role}.

In most tropical cyclone basins, satellite-based Dvorak technique or reconnaissance air-crafts are used to estimate MSWS \cite{windspeed}. These techniques provide reasonable estimates but require advanced machinery. Therefore, estimating MSWS from other tropical cyclone parameters is a significant problem. Much work has been done towards this problem; see \cite{Chaudhuri12Intensity, detailed} and references therein for a complete history of the work relating to cyclone intensity prediction.

We propose a method to estimate MSWS based on other characteristics of a tropical cyclone like date, time, latitude, longitude, pressure drop and estimated central pressure. We use machine learning algorithms to devise a regression model to estimate MSWS from other characteristics. We further employ machine learning classification algorithms to predict the grade of the cyclone based on these characteristics. 

\section{Materials and methods}
\subsection{Data}
The best track dataset of tropical cyclonic disturbances are collected from the Regional Specialized Meteorological Centre, New Delhi (\url{http://www.rsmcnewdelhi.imd.gov.in/index.php?option=com_content\&view=article\&id=48\&Itemid=194\&lang=en}) has been used in this study for the period from 1990 to 2017 in the North Indian ocean. The basin of origin, name (if there any), date and time of occurrence, position (latitude and longitude), Class number (or T No.), estimated central pressure, MSWS, pressure drop, grade, outermost closed isobar and diameter of outermost closed isobar of tropical cyclones are provided in the dataset. We define terms which we are going to use in the analysis below \cite{mssw}:

\begin{itemize}
    \item {\bf Basin of origin(BOO):} The Arabian sea, Bay of Bengal, or land is the possible basin of origins of any cyclone. 
    \item {\bf Date and Time:} The date and time of the origin of the cyclone.
    \item {\bf Latitude and Longitude:} The latitude and longitude in degrees along the path of the cyclone.
    \item {\bf Estimated central pressure (ECP):} It is the surface pressure at the center of the tropical cyclone as measured or estimated (in hPa (hectopascals)).
     \item {\bf Pressure drop (PD):} It is the drop in the pressure with respect to the atmospheric pressure. It is also measured in hPa. 
    \item {\bf Maximum sustained surface wind (MSWS):} The maximum sustained surface wind speed is the highest average of 3 minutes surface wind speed occurring within the circulation of the system. It is measured in knots (nautical miles per hour), which is the same as 1.86 Kilometers per hour.

    \item {\bf Grade:} Any tropical cyclone that develops within the North Indian Ocean between $100^{\circ}$E and $45^{\circ}$E is monitored by the India Meteorological Department (IMD). Tropical cyclone intensity scale according to cyclone category are given in the following table: 
    \begin{table}[!h]
    \centering
    \begin{tabular}{c|c|c} \hline
     \textbf{Grade} & \textbf{Low pressure system}  &  \textbf{MSWS (in knots) }\\ \hline
         1 & Low Pressure Area (LP) & $<$17\\
2 & Depression (D) & 17-27\\
3 & Deep Depression (DD) & 28-33\\ 
4 & Cyclonic Storm (CS) & 34-47\\ 
5 & Severe Cyclonic Storm (SCS) & 48-63\\ 
6 & Very Severe Cyclonic Storm (VSCS) & 64-119\\ 
7 & Super Cyclonic Storm (SS) & $\geq$120\\  \hline
    \end{tabular}
    \caption{The classification of the low pressure systems by IMD.}
    \label{tab:grade}
\end{table}
\end{itemize}

\begin{figure}[htbp!]
\centering
\includegraphics[height=12cm,width=12cm,keepaspectratio]{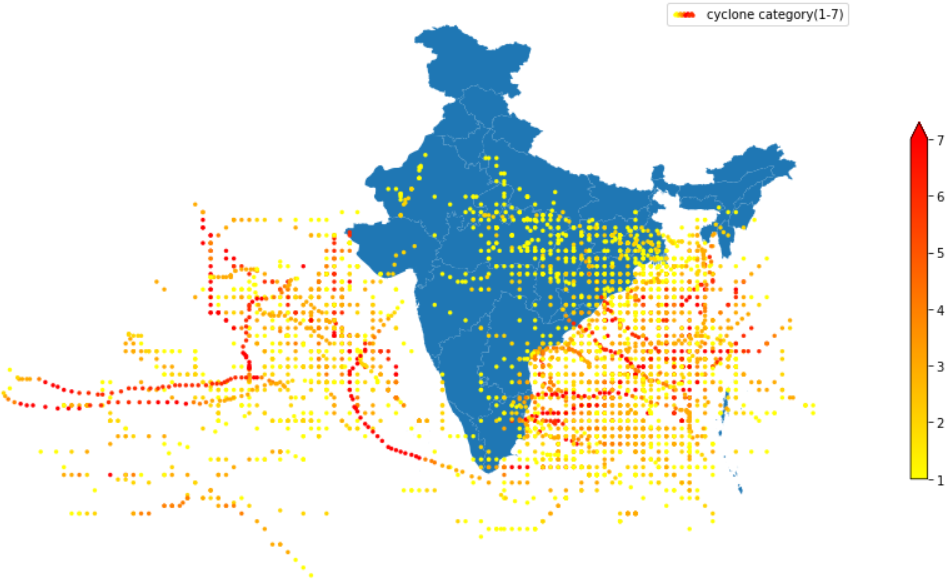}
\captionof{figure}{Cyclones hitting India since 1990-2017.}
\label{fig:india}
\end{figure}

There is a total of 4852 instances of cyclone measurements in the dataset used, out of which we selected 4021 for our study, dropping all of those instances which have any missing feature value. A pictorial description of these cyclones, along with colour-coded grade, is shown in Figure~\ref{fig:india}. The date is divided into three classes according to three seasons: Pre Monsoon - March to May, Monsoon- June to September, Post Monsoon - October to February. Time is divided into two categories according to day and night. We did not include outermost closed isobar and diameter of outermost closed isobar as attributes in our study, as very few data points were available in these columns. Table~\ref{tab:data} describes the distribution of data in different categories. 

\begin{table}[!htbp]
    \centering
    \begin{tabular}{c|c|c}
    \hline
   {\bf Characteristics} & {\bf Subdivisions} & {\bf Number of data points}\\ \hline 
        Basin of Origin & \begin{tabular}{@{}c@{}}Arabian Sea\\ Bay of Bengal \\ Land\end{tabular} & \begin{tabular}{@{}c@{}}
             1149 \\
             2707 \\
             165
        \end{tabular}\\ \hline
        Season & \begin{tabular}{@{}c@{}} 
            Pre-Monsoon (March - May)\\
            Monsoon (June to September)\\ 
            Post-monsoon (October - February)
        \end{tabular} &
         \begin{tabular}{@{}c@{}}
         
         791\\
         1208\\
         2022

        \end{tabular}\\ \hline
        Grade & \begin{tabular}{@{}c@{}}
             1  \\
             2 \\
             3 \\
             4 \\
             5 \\
             6 \\
             7 
        \end{tabular}     & \begin{tabular}{@{}c@{}}
             1433 \\
             915\\
             920\\
             266\\
             15\\
             205 \\
             267
        \end{tabular}\\ \hline
    \end{tabular}
    \caption{Baseline Data.}
    \label{tab:data}
\end{table}

\subsection{Methodology}

The MSWS is a continuous variable, while the grade is a categorical variable. Therefore, we use various machine learning regression and classification algorithms (XGBoost, Gradient Boosting Machine, Linear Regression, Decision Tree, Random Forest, SVM, Naive Bayes, Logistic Regression) for the prediction of MSWS and grade. In what follows, we briefly describe these algorithms.

\subsubsection{Decision tree} Decision Tree \cite{10.1023/A:1022643204877} is one of the most popular supervised machine learning algorithms used for both classification and regression techniques. The algorithm can be represented by an inverted tree with a root node at the top and other nodes connected to it through branches. Each node corresponds to a feature and a value assigned to the feature, while each branch represents a decision taken for the output variable based on the node it is emanating. To decide which feature to be placed at a node, we use measures like the Gini index, Entropy, or Information gain. For a given attribute $X$, 
\begin{itemize}
    \item Entropy is defined as
\[
E(X) = \sum -P(X=x)\log_2 (X=x),
\] 
\item Information gain is defined as
\[
IG(X,x)=E(X)-E(X|x),
\]
\item Gini index is defined as
\[
1-\sum (P(X=x))^2
\]
\end{itemize}
where $P$ denotes the probability, $E(X)$ denotes the entropy and $E(X|x)$ is the conditional entropy for a particular instance $x$ of $X$. We can determine the importance of a given attribute of a feature vector by calculating one of the above for that attribute. 

\subsubsection{Random Forest}
Random forest \cite{10.1023/A:1010933404324} is an ensemble learning method that can be used for both classification and regression. It generates multiple decision trees as part of the training process and outputs the mode (average) of these trees as per the classification (regression) problem. This approach solves the problem of overfitting, which is prevalent in the case of Decision Trees.

\subsubsection{Gradient Boosting Machine} 
Gradient Boosting Machine \cite{Friedman00greedyfunction} is an ensemble machine learning technique that is used for both classification and regression problems. It depends on the boosting technique where each weak learner is assigned a large weight to convert them to a strong learner in an iterative manner.

\subsubsection{XGBoost} 
\begin{figure}[!h]
\centering
\includegraphics[width=14cm,height=11cm,keepaspectratio]{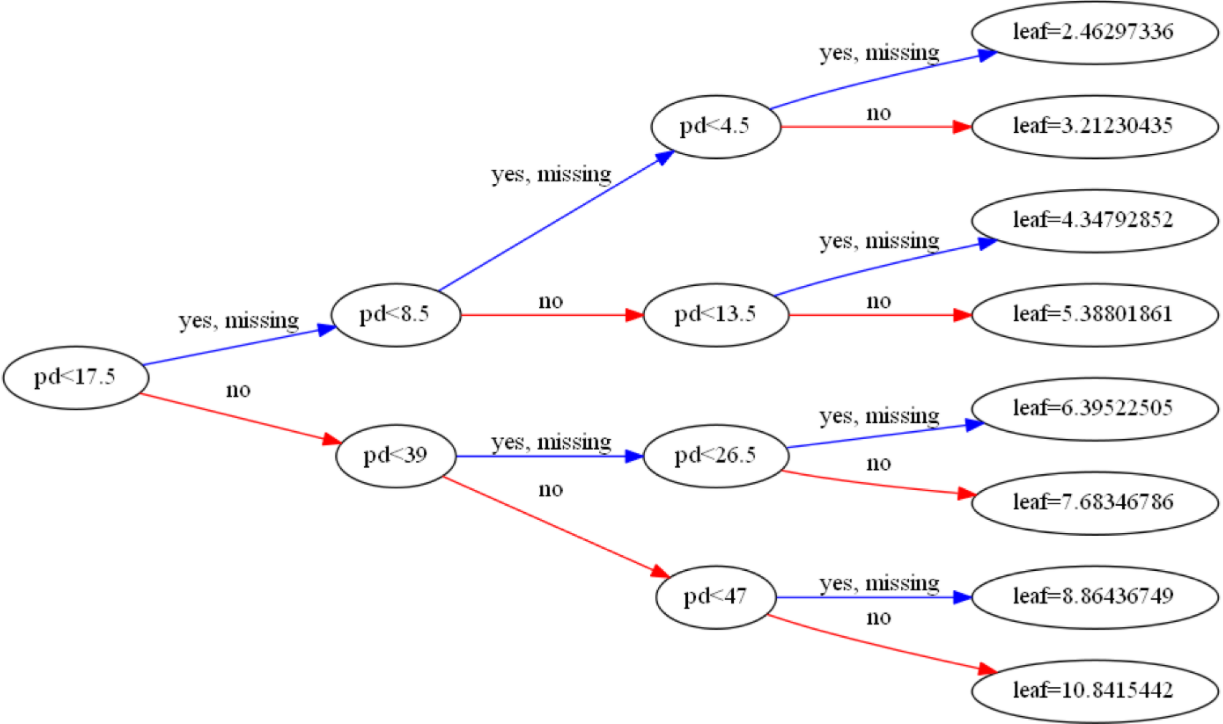}
\captionof{figure}{XGBoost tree for MSWS.}
\label{fig:XG}
\end{figure}

XGBoost \cite{DBLP:journals/corr/ChenG16} is one of the most popular recent supervised learning tree boosting scalable machine learning algorithms, which is based on function approximation and several regularization techniques. It is used for both classification and regression problems. Let $\widehat{ y_i}$ is the outcome from the ensemble model defined as follows:

\[
\widehat{ y_i}=\phi(x_i)=\sum_{k=1}^K f_k(x_i),\quad f_k\in \mathcal{F}
\]
where $\mathcal{F}=\{f(x)=w_{q(x)}\}$, $q:\mathbb{R}^m \rightarrow T$, $w\in \mathbb{R}^T$ is the space of all regression trees and $T$ denotes the total number of leaves in the tree.
In the above equation, $f_k$ represents a regression tree and $f_k(x_i)$ is the outcome given by the 
 $k$th tree to the $i$th entries in the data. The goal in XGBoost is to minimize the following regularized objective function: 
$$\mathcal{L}(\phi)= \sum_{i=1}^n l(y_i,\widehat{y_i})+\sum_{k=1}^K \Omega(f_k)$$
where $l$ is the loss function. To avoid high complexity of the model, a regularization term $\Omega$ is used which is given by
\[
\Omega(f_k)= \gamma T+\frac{1}{2}\lambda ||w||^2= \gamma T+\frac{1}{2}\lambda \sum_{j=1}^T w_j^2
\]
Where $\gamma$ and $\lambda$ are regularization parameters, the best split at any given node can be found from the following formula:
\[
\mathcal{L}_{split}=\frac{1}{2} \bigg[ \frac{G_L^2}{H_L+\lambda}+\frac{G_R^2}{H_R+\lambda}-\frac{(G_L+G_R)^2}{H_L+H_R+\lambda}\bigg]-\gamma
\]
Where $L$ stands for left-hand node and $R$ stands for right-hand node by letting $I=I_L\cup I_R$. Figure~\ref{fig:XG} shows the XGBoost tree for the estimation of MSWS.

\subsubsection{Linear Regression} In Linear Regression \cite{Jeffrey2001Linear}, a hyperplane is estimated that gives best linear relationship between independent variables (features) and dependent variable (target). The prediction model (hypothesis) is given by : 
\[
h_{\theta}(X) = \theta_0 + \theta_1 x_1 + \theta_2 x_2 + \dots + \theta_n x_n
\]
where $X = (x_1, x_2, \dots, x_n)$ represents the input vector and $\theta = (\theta_0, \theta_1, \theta_2, \dots, \theta_n)$ are the coefficients that determine the hyperplane. These coefficients are learned through an iterative process called gradient descent by minimizing the following loss function: 
\[
J(\theta) = \sum_{i=1}^{m} (h_{\theta}(X_i) - y_i)^2,
\] 
where $X_i$ denotes the $i$th input vector and $y_i$ corresponding target value.

\subsubsection{Logistic Regression} Logistic regression \cite{Walker1967Estimation} is a classifier that can be used to solve a multiclass prediction problem. Its an extension of Linear Regression, where the classification problem is converted into regression problem by estimating the log(odds) of each class in place of probability itself. If $p_i$ denotes the probability of $i$th class then the log(odds) for this class is defined as $p_i/\sum_{j\not = i}^{} p_j$. 

\subsubsection{Support Vector Machines (SVM)} SVM \cite{Cortes1995Support} can be used for both classification and regression problems. Like the Linear regression, SVM tries to find a separating hyperplane, but with maximum margin.  The learning problem is converted into an objective (nonlinear) maximization problem, subject to linear constraints. Using the tools of Linear Programming Problem (LPP), few input vectors (called support vectors) are selected that can be used for prediction. The nonlinear separating case of input vectors can be handled with kernels techniques.

\subsubsection{Naive Bayes} Naive Bayes \cite{Rish01anempirical} can be used for both classification and regression problems. The Naive Bayes algorithm is based on Bayes' theorem with an assumption that the features are linearly independent. Suppose $X_1, X_2, \cdots , X_n$ are real-valued attributes, and $Y$ is the set of all possible outcomes. Now according to the Bayes' theorem,
\[
P(Y=y_i|X_1,X_2,\cdots ,X_n)=\frac{P(Y=y_i)P(X_1,X_2,\cdots,X_n|Y=y_i)}{\sum_k P(Y=y_k)P(X_1,X_2,\cdots,X_n|Y=y_k)}.
\]
If we assume that $X_i$ are conditionally independent for given outcome set $Y$, then the above equation can be written as 
\[
P(Y=y_i|X_1,X_2,\cdots ,X_n)=\frac{P(Y=y_i)\prod_j P(X_j|Y=y_i)}{\sum_k P(Y=y_k)\prod_j P(X_j|Y=y_k)}
\]
The above equation is used for the classification problem. Similarly, we can define Naive Bayes for regression problems, where the sum in the above equation will be replaced by integration.

\subsubsection{Metrics}
To evaluate the performance of regression models for MSWS, we use the Root Mean square error (RMSE) and Coefficient of determination ($R^2$). 
\begin{itemize}
    \item \textbf{RMSE:} If there are $m$ sample points with $y_i$ as actual value and $\widehat{y_i}$ as predicted value evaluated from the model, then RMSE is defined as 
\[
\sqrt{\frac{1}{m}\sum_{i=1}^{m}(y_i-\widehat{y_i})^2}.
\] 
RMSE is always non-negative and should be close to $0$.

\item \textbf{$R^2$:} The coefficient of determination ($R^2$) is defined as 
\[
R^2 = 1- \frac{\text{SE}_{\hat{y}}}{\text{SE}_{\bar{y}}}
\]
where total sum of squares, $\text{SE}_{\bar{y}}$, is defined as $\text{SE}_{\bar{y}}=  \sum_{i=1}^{m}(y_i-\bar{y_i})^2$ and residual sum of squares, $\text{SE}_{\hat{y}}$ is defined as 
$\text{SE}_{\hat{y}}=\sum_{i=1}^{m}(y_i-\widehat{y_i})^2$. Here, $\bar{y}$ is the mean of the data, $\bar{y}= \frac{1}{m} \sum_{i=1}^{m}y_i$.
\end{itemize}

The confusion matrix is used to determine the performance of the classification model on the test data. For classification models, multi-class classification accuracy has been measured using the confusion matrix \cite{confusionmatrix}. Accuracy is the ratio between all correctly predicted samples to all possible samples.
\[
     \text{Accuracy} = \frac{\text{correctly predicted samples}}{\text{total number of test samples}}
\]

\section{Results and Discussions}\label{sec:results}

\subsection{Correlation analysis}
\begin{figure}[!htbp]
\centering
\includegraphics[width=14cm,height=8.5cm,keepaspectratio]{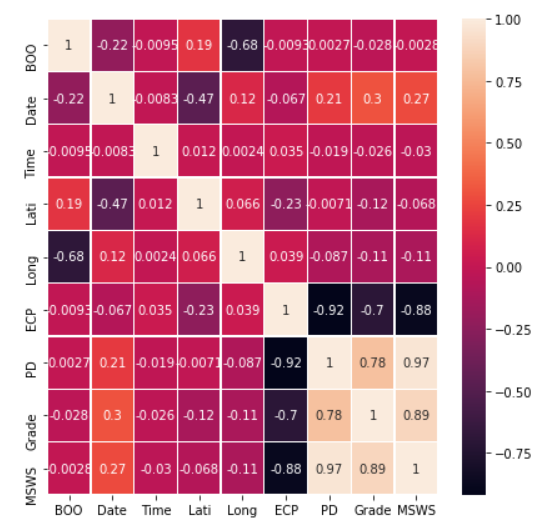}
\captionof{figure}{Correlation between variables.}
\label{tab:corr}
\end{figure}
The correlation matrix of all variables is given in Figure~\ref{tab:corr}. The grade is weakly correlated with all the variables except ECP and PD. Also, the correlation of grade with ECP is negative, suggesting that if central pressure is low, the intensity of the cyclone is high. The MSWS shares a similar correlation with ECP as the grade. This is not surprising as grade is directly evaluated from MSWS; see Table~\ref{tab:grade}. PD has a strong positive correlation with MSWS. A linear regression suggests the following relationship between MSWS and PD
\[
\operatorname{MSWS} \approx 1.6 \operatorname{PD} + 22.3
\]
in the North Indian Ocean. Notice that in \cite{Rosendal982Relationship}, a similar relationship between MSWS and PD ($\text{MSWS} \approx 1.176 \text{PD} + 30$) was reported for tropical cyclones in Central North Pacific Ocean.

\subsection{Model selection and validation}
We use 10-fold cross-validation for each of the models. In each fold, we split the data into training and validation sets in the ratio of 4:1. Then, each ML algorithm is applied to the training set to train the model. At every step, the performances (RMSE, $R^2$, or accuracy) of the model are recorded, and the average of each of these performances is reported in Tables~\ref{tab:mssw} and \ref{tab:grade1}.      


\begin{table}[!h] 
\begin{subtable}{.48\textwidth}
\centering    
\begin{tabular}{ c|c|c } 
\hline
 Model  &  RMSE & $R^2$ \\
\hline
XGBoost & 2.30 & .99 \\
\hline
 \makecell{Gradient Boosting\\ Machine} & 2.80 & 0.97 \\
\hline
Decision Tree & 3.91 & 0.94 \\
\hline
Random Forest & 3.12 & 0.96 \\
\hline

Linear Regression & 5.07 & 0.92 \\
\hline
\makecell{SVM\\ Kernel:-RBF\\ Kernel:Linear\\ Kernel: Polynomial\\ (4th-degree)} & \makecell{6.11\\ 5.69\\ 3.93} & \makecell{0.90\\ 0.91\\ 0.95} \\
\hline
Naive Bayes & 3.38 & 0.97 \\
\hline 
\end{tabular}
\caption{Regression Analysis on MSWS.}
 \label{tab:mssw}
 \end{subtable}
 \hfill
  \begin{subtable}{.48\textwidth}
  \centering    
\begin{tabular}{ c|c } 
\hline
Model & Accuracy\\
\hline 
XGBoost & 87.15 \\
\hline 
GBM & 85.73 \\
\hline 
\makecell{Decision Tree\\Entropy(Depth-4)\\ Gini(Depth-4)} & \makecell{87.91 \\ 84.76} \\
\hline
Random forest & 85.95 \\
\hline
Naive Bayes & 86.39 \\
\hline 
Logistic & 71.28 \\
\hline 
\makecell{SVM\\ Kernel:-RBF\\ Kernel: Linear\\ Kernel: Polynomial\\ (degree 4)} & \makecell{78.48\\ 86.90\\ 84.36} \\
\hline
\end{tabular}
\caption{Classification(Multi-class) Analysis on Cyclone grade.}
 \label{tab:grade1}
  \end{subtable}
 \end{table}
It is evident from Table~\ref{tab:mssw} that XGBoost is outperforming other models with an RMSE of 2.3 and $R^2$ of 0.99. Notice from Table~\ref{tab:grade} that the range of values of MSWS for a particular grade is always greater than or equal to 5, and since XGBoost is predicting MSWS with an RMSE of 2.3, we expect that XGBoost will also predict grade with very high accuracy. That is definitely the case, as from Table~\ref{tab:grade1}, XGboost has an accuracy of 87.15\% in predicting the grade. However, the Decision Tree with Entropy of depth 4 outperforms XGBoost in predicting the grade with an accuracy of 87.91\%. 

Moreover, if we fix the classification model for the grade to be the Decision Tree with Entropy of depth 4, Table~\ref{tab:cat} represents the accuracy in predicting a particular category for the grade. The model predicts the top three high-intensity categories (SCS, VSCS, and SS) of grade with an average accuracy of 98.84\%. 

\begin{table}[!htbp]
\centering    
\begin{tabular}{ c|c } 
\hline
Category & Accuracy\\
\hline
LP & 98.33\\
\hline
D & 77.92\\
\hline
DD & 78.37\\
\hline
CS & 88.72\\
\hline
SCS & 100\\
\hline
VSCS & 99.51 \\
\hline
SS & 97 \\
\hline
\end{tabular}
\caption{Classification accuracy of different Cyclone Grade.}
\label{tab:cat}
 \end{table}

\subsection{Testing on Vayu and Fani}
\begin{figure}[!h]
\centering
\begin{minipage}{.5\textwidth}
   \centering
  \hbox{\hspace{-0.7cm}\includegraphics[width=0.97\linewidth]{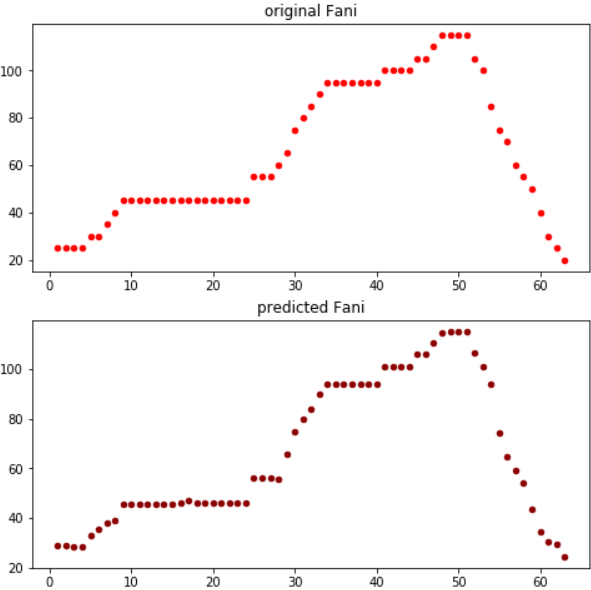}}
\end{minipage}%
\begin{minipage}{.5\textwidth}
  \centering
  \hbox{\hspace{0.4cm}\includegraphics[width=0.97\linewidth]{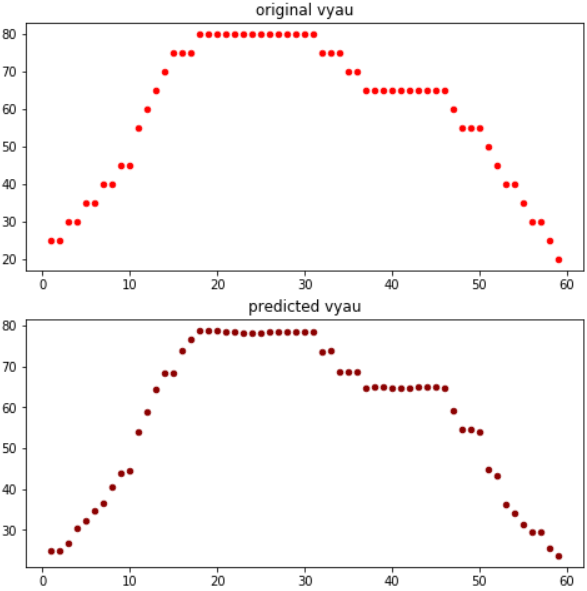}}
\end{minipage}
\captionof{figure}{Scatter plot of actual and model-predicted MSWS for Fani and Vayu.}
\label{fig:vayufanimssw}
\end{figure}

\begin{figure}[!h]
\centering
\begin{minipage}{.48\textwidth}
  \centering
 \hbox{\hspace{-1cm}\includegraphics[width=0.97\linewidth]{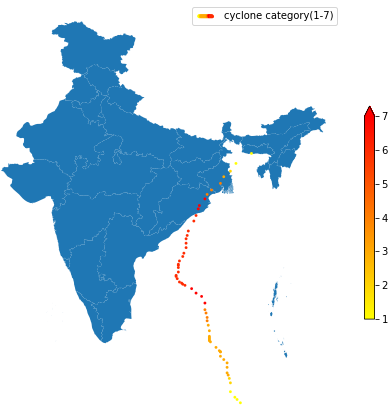}}
\end{minipage}%
\begin{minipage}{.48\textwidth}
  \centering
  \hbox{\hspace{0.4cm}\includegraphics[width=0.97\linewidth]{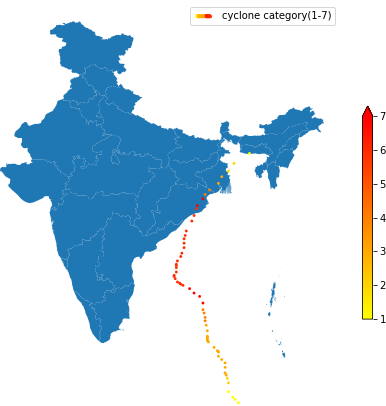}}
\end{minipage}
\captionof{figure}{Actual and model predicted grade along track of Fani.}
\label{fig:fanigrade}
  \end{figure}

\begin{figure}[!h]
\centering
\begin{minipage}{.48\textwidth}
  \centering
 \hbox{\hspace{-0.8cm}\includegraphics[width=0.97\linewidth]{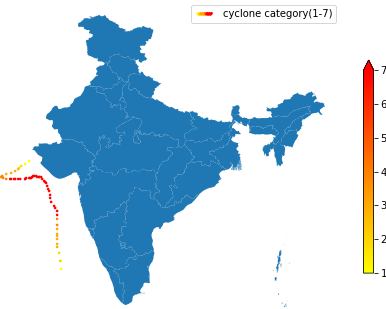}}
\end{minipage}%
\begin{minipage}{.48\textwidth}
  \centering
  \hbox{\hspace{0.5cm}\includegraphics[width=0.97\linewidth]{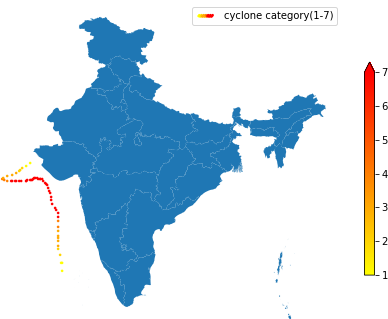}}
\end{minipage}
\captionof{figure}{Actual and model predicted grade along track of Vayu.}
\label{fig:vayugrade}
\end{figure}
We test our model on two recent tropical cyclones, Vayu and Fani. Vayu was a grade 7 tropical cyclone which hit the Indian west coast in June 2019. Around 6.6 million people were affected in northwestern India by the cyclone \cite{VAYU}. Fani was also a grade 7 tropical cyclone that hit the Indian state of Odisha in April-May 2019. Due to Fani, India and Bangladesh faced heavy damages. At least 89 people have been reported died, and damages caused estimated around US\$8.1 billion \cite{FANI}. 

We checked the performance of the best model to predict MSWS, XGBoost, on Vayu and Fani. The RMSE is 2.2 and 3.4, while $R^2$ is 0.99 and 0.99 for Vayu and Fani, respectively. Figure~\ref{fig:vayufanimssw} depicts the actual values of MSWS and values predicted by the XGBoost model during the course of Vayu and Fani.

The best model to predict grade, Decision Tree with Entropy with depth 4, predicts different grades during the course of Vayu and Fani with an accuracy of 93.22\% and 95.23\%, respectively. The actual and predicted grades along the track of Vayu and Fani is shown in Figures~\ref{fig:fanigrade} and \ref{fig:vayugrade}.

\section{Conclusion} Estimating the intensity of tropical cyclones on a real-time basis is a problem worth studying, considering the human life and economic loss involved. In this study, we explored various machine learning techniques and reported their performance to estimate the Maximum Surface Sustained Wind Speed and intensity of the tropical cyclone. Our research finds that the ML model XGBoost and Decision Tree can be used for the estimation of MSWS and intensity with excellent performance over the North Indian ocean.

\section*{Acknowledgement} Authors are thankful to the Indian Meteorological Department (IMD) for providing the data archives.

\section*{Conflict of Interest} All the authors declare that they have no conflict of interest.



\bibliographystyle{unsrt} 
\bibliography{reference.bib}

\end{document}